\documentclass[twocolumn]{aastex631}

\usepackage{CJK}
\usepackage{multirow}

\begin{document}
\begin{CJK*}{UTF8}{gbsn}


\title{Fast inflowing ionized absorber tracing the gas dynamics at sub-parsec scale around Mrk\,3}

\correspondingauthor{Fangzheng Shi}
\email{fzshi@shao.ac.cn}

\author[0000-0003-3922-5007]{Fangzheng Shi (施方正)}
\affiliation{Shanghai Astronomical Observatory, Chinese Academy of Science, No.80 Nandan Road, Shanghai, China}

\author{Matteo Guainazzi}
\affiliation{European Space Agency (ESA), European Space Research and Technology Centre (ESTEC), Noordwijk, The Netherlands}

\author{Yijun Wang (王倚君)}
\affiliation{School of Astronomy and Space Science, Nanjing University, 163 Xianlin Avenue, Nanjing 210023, People’s Republic of China}
\affiliation{Key Laboratory of Modern Astronomy and Astrophysics, Nanjing University, Ministry of Education, 163 Xianlin Avenue, Nanjing 210023, People’s Republic of China}





\begin{abstract}
Accretion onto supermassive black hole (SMBH) can release energy via radiation, jets or winds, providing feedback effects on the circumnuclear gas environment.
However,
not all active galactic nuclei (AGNs) exhibit clear signature of such feedback, and
the dynamics of accreting gas on the inner sub-parsec scales remains poorly understood.
Using high-resolution Chandra X-ray grating spectra of Mrk\,3, we detect a fast inflowing ionized absorber characterized by redshifted Fe XXV and Fe XXVI absorption lines with confidence level in the $94-99.6\%$ range. 
Photoionization modeling reveals the inflowing absorber is located at $\lesssim0.04-0.74\rm~pc$ , with red-shifted velocity decreasing from $6.1\pm0.5\times10^3\rm~km~s^{-1}$ to $3.4\pm0.3\times10^3\rm~km~s^{-1}$ over 11 years.
Only $\sim0.6$\%--$3$\% of the inflowing material is estimated to reach the event horizon.
This direct evidence of sub-parsec scale fueling inflow bridges the gap between the torus and the outer accretion disk.
Additionally, a $0.86$-keV gas component with sub-solar metallicity ($Z\sim0.22$), outflowing at a velocity of $\sim330\rm~km~s^{-1}$, is detected in the soft X-ray band from XMM-Newton Reflection Grating Spectrometer, probably corresponding to shocked interstellar medium in the narrow-line region (NLR).
The simultaneous presence of the apparent decelerating sub-parsec inflow and the NLR outflow favors a coherent scenario where a putative disk wind or broad-line region clouds may impede or even eject the accretion material, although other possibilities cannot be fully excluded.
\end{abstract}

\keywords{High energy astrophysics (739) --- Accretion(14) --- X-ray active galactic nuclei(2035) --- Photoionization(2060)}


\section{Introduction} \label{sec:intro}

A supermassive black hole (SMBH) at the center of nearly every galaxy has been considered as an essential component to galaxy evolution.
Actively accreting SMBHs, manifesting themselves as active galactic nuclei (AGN) can convert accretion power into considerable energy output in the form of radiation, jets or winds, thereby regulating
both black hole growth and star formation--an effect known as AGN feedback \citep{2012ARA&A..50..455F}.
The infalling motion of cold accretion gas toward the central SMBH can be traced from extragalactic scales (through galaxy mergers and interaction), galactic scales (through secular processes), down to sub-hundred-parsec scale (through nuclear gas disks, bars or spirals) \citep{2019NatAs...3...48S}.
At smaller scales, \citet{2016ApJ...829...96S} detected a redshifted absorber in the spectra of a quasar and found that the inflowing medium is located at $\sim4$ pc from the black hole near the AGN torus with reverberation mapping.

However, observational evidence of cold gas inflow that directly feeds black hole accretion disks on sub-parsec scales remains limited.
In a small sample of quasars around $z\sim1$ (e.g, in J1035+1422), redshifted optical broad absorption lines reaching $\sim1000-5000\rm~km~s^{-1}$ have been detected coexisting with blueshifted absorption lines.
These observations features a potential feeding inflow, possibly connecting between the torus and the accretion disk within several $10^{3-4}\rm~r_{g}$, in addition to the disk wind \citep{2019Natur.573...83Z,2022A&A...659A.103C}.
\citet{2023Sci...382..554I} resolved a gravitationally unstable, dense gas disk driving sub-parsec scale accretion inflow onto the black hole in the Circinus Galaxy. Only 3\% of the inflowing material is estimated to reach the event horizon, with the rest ejected by multiphase outflows.

\citet{2019Natur.573...83Z} proposes the AGN torus as the gas reservoir fueling the accretion disk and attributes the scarcity of observed disk-feeding inflow to an observational bias due to line-of-sight effect. When the line of sight is close to the polar direction, it intercepts only the blueshifted, disk driven outflow aligned with the narrow-line region (NLR). Conversely, when the line of sight inclining towards the torus plane, the inflow material would be totally obscured by the optically thick torus.
Only a modest inclination angle grazing the surface of the dusty torus, would enable a direct detection piercing through the inflowing medium connecting to the accretion disk.

As gas approaches sub-parsec scale toward the central AGN, it becomes heavily photoionized by the intense radiation field.
The physical properties of such highly ionized gas are best probed through X-ray spectroscopy.
Highly ionized, ultra-fast outflows, with velocities up to 0.6$c$, characterized by blueshifted UV / X-ray absorption lines are ubiquitously found in up to $30$--$50$\% of local AGN \citep{2012ApJ...753...75C,2010A&A...521A..57T,2021ApJ...920...24C}, while their redshifted counterparts,
fast inflows at sub-parsec scales are rarely reported.
Highly ionized inflows have also been detected in the central AGN of a limited sample, including nearby Seyfert 1 galaxies\citep{2005A&A...442..461D,2007MNRAS.374..237L,2017A&A...597A..66G}, quasars\citep{2005ApJ...633L..81R, 2018MNRAS.481.1832P} and the Seyfert 2 galaxy ESP 39607 \citep{2025arXiv250507963P}, with velocities in the range of $0.15$ -- $0.4\rm~c$, much higher than those detected with optical absorption lines. This suggests that these absorbers are located very close to the central black hole, on a scale of accretion disk (several to several hundreds gravitational radii $r_{\rm g}$). These features have been interpreted as signatures of matter directly falling onto the black hole in the innermost part of accretion disk.
Still evidence for disk-feeding inflow remains scarce.
Therefore, detailed X-ray spectroscopic studies of individual AGN observed at grazing angles in the local Universe would help reveal the complex `inflow-outflow' gas dynamics in the very vicinity of AGN and the narrow-line region.

Lying at a distance of $58.5\rm~Mpc$ ($z\sim0.0135$\footnote{\url{https://ned.ipac.caltech.edu}}), Mrk\,3 (Markarian 3) is an {\it S0}-type galaxy hosting a nearly obscured AGN\citep{2015MNRAS.454..973Y}.
The estimated angle-averaged column density of the obscuring medium over all directions is below the threshold of $N_{\rm H}\lesssim1.24\times10^{24}\rm~cm^{-2}$, corresponding to a Compton optical depth of $\tau_{\rm S}\lesssim1$\citep{2015MNRAS.454..973Y,2016MNRAS.460.1954G}.
Thus it is classified as a Compton-thin AGN.
The mass of the central supermassive black hole is estimated to be $log(M_{\rm BH}/\rm M_{\rm\odot})=8.65$ based on stellar velocity dispersion \citep{2002ApJ...579..530W}.
The Eddington ratio of Mrk\,3 ranges between $0.01$ and $0.04$.
Signatures of ionized gas outflows have been discovered and well-studied in Mrk\,3.
Kinematic modeling of emission lines (e.g., [O III], [N II]) reveals a peak gas velocity of $\sim500-1500\rm~km~s^{-1}$ in the narrow-line region (NLR), tracing a biconical outflow extending to $\sim400\rm~pc$ \citep{2001AJ....122.2961R, 2005ApJ...619..116C, 2020ApJ...893...80G, 2021ApJ...910..139R,2023ApJ...943...98M}.
Diffuse X-ray emission observed in Chandra non-grating image reveals that the highly ionized gas extends along the direction roughly coincident with the [O III] outflow in the NLR \citep{2017ApJ...848...61B}.
\citet{2016MNRAS.460.1954G} analyzed the X-ray broadband spectrum of Mrk\,3 with the state-of-the-art torus model to investigate its torus properties (including the torus opening angle). 
They concluded that the line-of-sight intercepts the torus surface. 
This makes Mrk\,3 analogous to those $z\sim1-2$ quasars with a torus-grazing viewing angle mentioned above.
They also identified H-like and He-like iron absorption lines in the stacked Chandra grating spectrum, indicating the presence of a potential ionized absorber. However, they did not constrain the physical properties of this absorber and left its origin unexplained.

Our motivation is to investigate the properties of the ionized absorber in detail, to explore the connection between torus, accretion inflows and outflows, and to get a better understanding of the hot gas dynamics and feedback effect in the vicinity of the central AGN in Mrk\,3.
This paper is organized as follows. In Section \ref{sec:reduc} we describe the data reduction processes in detail. In Section \ref{sec:res}, results of spectral fitting for both soft and hard X-ray bands are presented, as well as a detailed photoionization modeling for potential redshifted ionized absorber.
In Section \ref{sec:diss}, we try to propose a preferred scenario to explain the multi-scale, multi-phase dynamical behavior of the gas from sub-parsec to hundred parsec around Mrk\,3 nuclei.
We summarize our findings in Section \ref{sec:sum}.

\section{Data Reduction} \label{sec:reduc}
To detect plasma with bulk motion around the Mrk\,3 nucleus, all available high-resolution X-ray spectroscopic data covering both soft (6.2--24.6 $\rm\AA$ or 0.5--2 keV) and hard (1.6--6.2 $\rm\AA$ or 2--10 keV) X-ray bands has been retrieved.
The High Energy Transmission Grating (HETG) onboard Chandra X-ray Observatory, covering the energy band ranging from $1$ -- $10\rm~keV$, can achieve a spectral resolution of $\sim2000\rm~km~s^{-1}$ around $6$ -- $7.5\rm~keV$.
For the soft X-ray band, the XMM-Newton Reflection Grating Spectrometer (RGS) offers a first-order full-width half maximum (FWHM) spectral resolution of $\sim1200\rm~km~s^{-1}$ over $0.5$--$2\rm~keV$ range.
We summarize the basic information of all the observations in Table \ref{tab:obsid}.

\subsection{HETG} \label{subsec:hetg}
Mrk\,3 has been observed by HETG in 2000 and 2011 for a total clean exposure of $389.3\rm~ks$.
The data employed in this paper can be obtained by the Chandra X-ray Observatory, contained in the Chandra Data Collection 412~\dataset[DOI:10.25574/cdc.412]{https://doi.org/10.25574/cdc.412}.
We reduce the raw data with the CIAO v4.15 software package \citep{2006SPIE.6270E..1VF}, \texttt{chandra\_repro} with the calibration database CALDB v4.12.0.
The $\pm1st$-order High Energy Gratings (HEG) spectra have been extracted for all epochs by \texttt{tgextract} from the clean event files with default source and background extraction width.
Half-width of the source extraction region in the cross-dispersion direction is $2.39\rm\arcsec$, corresponding to $\lesssim680\rm~pc$ in physical size for Mrk\,3.
The underlying continuum shows little variability,
so all HEG spectra are co-added for a baseline fit with \texttt{combine\_grating\_spectra} (See Sec.\ref{subsec:base} for further details). While the line features are analyzed separately for each epoch.
It is worth noting that HETG observations used in this work are identical to those analyzed in \citet{2016MNRAS.460.1954G} with the same ObsID while they reported a total exposure time of $778.6\rm~ks$. The apparent discrepancy arises from different spectral stacking methods. The exposure time for +1 and -1 order of each epoch were separately added and counted twice in their work, while our method only counts once based on each event file.

\subsection{RGS} \label{subsec:rgs}
We retrieve 15 epochs of RGS observations for Mrk\,3.
After inspecting the background light curve,  exposure time with high flaring particle background has been removed, with a threshold of $0.2\rm~cts~s^{-1}$ for observations after 2012, $0.3\rm~cts~s^{-1}$ for ObsID 0009221601 and $0.1\rm~cts~s^{-1}$ for the rest of observations.
Four epochs with an all-time high background have been abandoned, leaving a total clean exposure of $130.8\rm~ks$ as seen in Table \ref{tab:obsid}.
We extract the spectra from the two RGS modules following the standard pipeline {\it rgsproc} with the SAS v21.0.0 software package \citep{2004ASPC..314..759G}. 
We check the individual first order spectra for each epoch and find very little spectral variability across all 11 selected epochs.
Thus all the first order RGS spectra are co-added together with \texttt{rgscombine}.

\begin{deluxetable}{lccc}
\tablecaption{Log of X-ray observations}
\label{tab:obsid}
\tablewidth{0pt}
\tablehead{
\colhead{Inst} & \colhead{Obs ID} & \colhead{Start Date} & \colhead{Net Exposure} \\
\colhead{  }   & \colhead{   }  & \colhead{(UT)}       & \colhead{(ks)}
}
\startdata
HETG & 873    & 2000-03-18 06:58:08 & 100.6        \\
HETG & 12874  & 2011-04-19 15:53:47 & 77.06        \\
HETG & 12875  & 2011-04-25 09:34:32 & 29.86        \\
HETG & 13254  & 2011-08-26 02:56:58 & 31.53        \\
HETG & 13261  & 2011-05-02 00:17:30 & 22.08        \\
HETG & 13263  & 2011-04-28 23:08:33 & 19.72        \\
HETG & 13264  & 2011-04-27 19:45:10 & 35.76        \\
HETG & 13406  & 2011-05-03 12:54:21 & 21.43        \\
HETG & 14331  & 2011-08-28 13:09:38 & 51.21        \\
RGS & 0009220301 & 2001-03-12 23:10:35 & 4.2 \\
RGS & 0009220401 & 2002-03-10 13:17:59 & 16.9 \\
RGS & 0009220501 & 2002-03-25 17:05:48 & 13.6 \\
RGS & 0009220601 & 2001-03-20 20:03:15 & 22.5 \\
RGS & 0009220701 & 2001-03-28 20:47:56 & 6.9 \\
RGS & 0009220901 & 2001-09-12 00:27:09 & 5.1 \\
RGS & 0009221601 & 2002-09-16 05:09:12 & 4.5 \\
RGS & 0656580301 & 2012-03-15 11:27:22 & 40.8 \\
RGS & 0741050101 & 2015-03-19 17:32:26 & 4.0 \\
RGS & 0741050201 & 2015-04-08 16:23:36 & 5.8 \\
RGS & 0741050401 & 2015-04-20 15:39:41 & 6.5 \\
\enddata
\end{deluxetable}

\section{Results} \label{sec:res}
Spectra have been optimally rebinned and analyzed with SPEX v 3.08.01 \citep{kaastra_2024_spex}.
We adopt the Cash statistics (C-stat) \citep{1979ApJ...228..939C} as the primary criterion for determining the best-fit models. In specific cases, we also use the corrected Akaike Information Criterion (cAIC) as a supplementary metric to assist in model comparison (see Section \ref{subsec:softX} for calculation details).
Galactic foreground absorption with column density $N_{\rm H,gal}=9.28\times10^{22}\rm~cm^{-2}$ has always been included in spectral modeling by introducing a \texttt{hot} model with temperature fixed at the lower limit of $92.8\rm~K$ to account for absorption from neutral HI\citep{2016A&A...594A.116H}.
Error bars of the derived properties are taken as the 1-$\sigma$ (68.3\%) confidence level, unless otherwise mentioned in the following sections.

\subsection{Modeling Baseline Continuum} \label{subsec:base}

Co-added HEG spectrum of Mrk\,3 over the 1.6--6.2 $\rm\AA$ (2--10 keV) band consists of an intrinsically absorbed power law with a distinct absorption edge around $7\rm~keV$, a prominent fluorescent Fe K$\alpha$ line and a reflection hump.
We adopt a photoionized absorber (\texttt{pion}(abs)$_{\rm torus}$) in front of the baseline power-law (\texttt{pow}) to account for the intrinsic absorption.
\texttt{SPEX} can jointly optimize the illuminating continuum and the \texttt{PION} parameters in a self-consistent way.
A reflection component (\texttt{refl}) has also been added to describe the baseline continuum.
The absorption edge energy ($\simeq7.11\rm~keV$ in the source frame) indicates that the intrinsic absorber is almost neutral. 
The ionization parameter $\xi$ is defined as $\frac{L}{nr^2}$, where $L$ represents the ionizing luminosity ($13.6\rm~eV$--$13.6\rm~keV$) of the source, $n$ denotes the hydrogen number density and $r$ gives the distance of the illuminated material.
Allowing the ionization parameter of {\it pion$_{\rm torus}$} to vary freely during the fitting causes it peg to the hard lower limit permitted by the model. Furthermore, we find that the goodness of fit ($C$-stat) and the best-fit value of other parameters are insensitive to $\xi$ for log($\xi$)\textless$-2$.
So we fix log($\xi$) of this component to the 3-$\sigma$ upper limit error range of $-2$.
The best-fit column density of this neutral absorber is $1.73_{-0.13}^{+0.14}\times10^{24}\rm~cm^{-2}$ and it could be associated with the torus around the AGN.
The reflection fraction, denoted as $scal = \frac{I_{\rm refl}}{I_{\rm PL}}$, is defined as the intensity ratio between the reflection component $I_{\rm refl}$ and the unabsorbed power-law component $I_{\rm PL}$. 
The best-fit $scal$ for the co-added spectrum is $0.20^{+0.07}_{-0.05}$.
Other best-fit parameters have been recorded in the column $\rm HEG_{con}$ of Table \ref{tab:best_fit_HEG}.

Our result is roughly consistent with previous studies, indicating that the intrinsic torus column density in Mrk\,3 lies near the boundary between the Compton-thin and Compton-thick regimes.
\citet{2015MNRAS.454..973Y} and \citet{2016MNRAS.460.1954G} adopt self-consistent torus model with careful analysis, and calculate the angle-averaged column density of the torus. They
conclude that the central AGN should be classified as Compton-thin.
It should be noted that there are slight discrepancies between our derived torus column density and previous studies.
The main reason is that,
we treated the absorbed and scattered continuum separately with \texttt{pion} and \texttt{refl} models.
Furthermore, the reflection model (\texttt{refl}) we applied in {\it SPEX} is built for plane-parallel, Compton-thick scattering material. 
This assumption is over simplified as the azimuthal column density distribution of the torus is not constant proved by \citet{2016MNRAS.460.1954G}.
Our baseline model is therefore not fully self-consistent in this context.
However, this approximation does not significantly affect any of the main conclusions of our analysis of the ionized inflow in this source.

We also divide spectra of all the epochs into 2 groups depending on the observation time (epoch 2000 and 2011). 
The same set of continuum models as used for the co-added HEG spectrum is adopted to the two individual groups of spectra. 
The unabsorbed 2-10 keV luminosity of the incident power law component increases from $5\times10^{44}\rm~erg~s^{-1}$ to $7\times10^{44}\rm~erg~s^{-1}$,
accompanied by a spectral softening. The photon index increases from $2.07^{+0.07}_{-0.08}$ to $2.32^{+0.05}_{-0.06}$.
A detailed analysis of the residual spectral features will be discussed in the following Section \ref{subsec:redWA}.
The best-fit results for the two groups are recorded in Table \ref{tab:best_fit_HEG} and Figure \ref{fig:heg_epoch}.

\subsection{redshifted Ionized Absorber} \label{subsec:redWA}
\subsubsection{Identification of absorption lines}
\label{subsubsec:line_identi}
Two residual absorption line features can be found around $6.5$--$6.8\rm~keV$ in the co-added HEG spectrum.
First we adopted two additional Gaussian models to the continuum model.
After correcting for the systematic redshift of Mrk\,3,
the centroids of these two absorption lines are $6.60\pm0.04\rm~keV$ and $6.82\pm0.03\rm~keV$.
The separation between their line centroids are consistent with that between the resonant He-like (rest-frame at $6.70\rm~keV$) and H-like (rest-frame at $6.97\rm~keV$) Fe K$\alpha$ lines within the error range.
If the detected lines are assigned to them, the Doppler redshifts of the two absorption line centroids are $4.5\pm1.8\times10^3\rm~km~s^{-1}$ and $6.5\pm1.3\times10^3\rm~km~s^{-1}$ with respect to their rest-frame energy,
consistent with each other within the error range. The Doppler velocity is calculated by $\sim\frac{E_{\rm rf}-E_{\rm obs}}{E_{\rm rf}}c$, where the $E_{\rm rf}$ and $E_{\rm obs}$ represent the rest-frame and observed line centroids, and $c$ denotes the speed of light.

\begin{figure}[hbtp!]
\centering\includegraphics[width=0.45\textwidth]{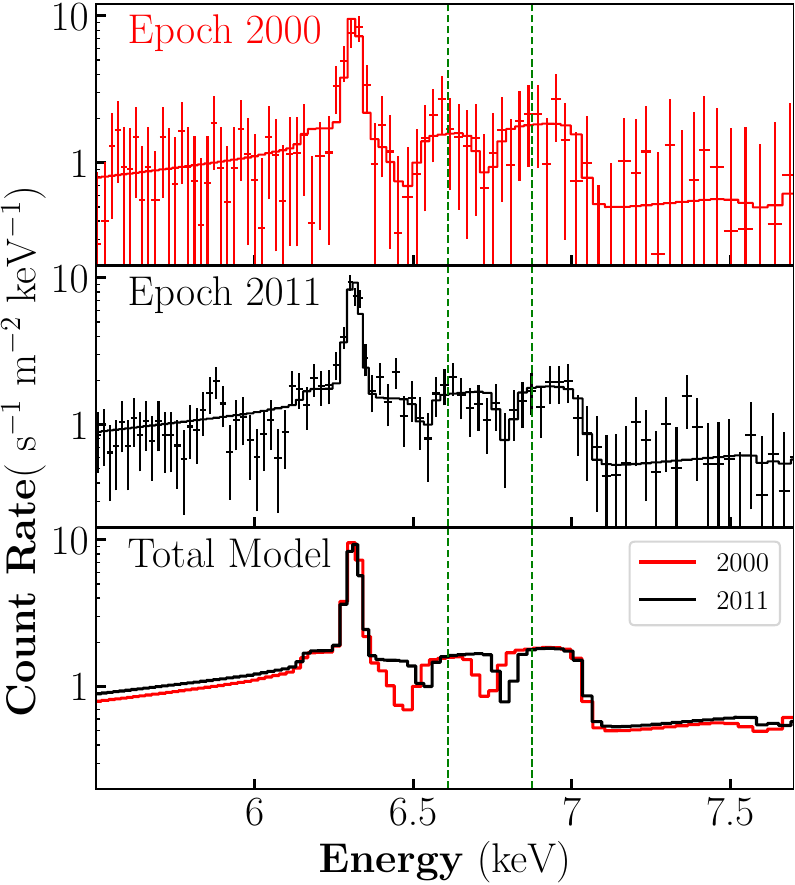}
\caption{
{\it The top and middle panel}: HEG spectra of Mrk\,3 for epoch 2000 (red) and epoch 2011 (black).
A redshifted photoionized absorber has been added to the baseline continuum model.
Crosses mark the observation data and solid lines represent the best-fit model.
Green dashed vertical lines indicate the rest-frame centroids for H-like ($6.70\rm~keV$) and He-like ($6.97\rm~keV$) iron lines in Mrk\,3. These absorption features are persistent over a time interval spanning over 11 yrs, while the inflowing velocity slightly decreases from $6.1\pm0.5\times10^3\rm~km~s^{-1}$ to $3.4\pm0.3\times10^3\rm~km~s^{-1}$. The column density and ionization parameter increase over the time.
{\it The bottom panel}: Comparison of the best-fit total models between epoch 2000 (red line) and epoch 2011 (black line).
Clear shifts of absorption line centroids are present across epochs.
}
\label{fig:heg_epoch}
\end{figure}

We further noticed that the two absorption-line features are present in both 2000 and 2011 observations (See Fig. \ref{fig:heg_epoch}) and exhibit clear variability across epochs.
Phenomenological Gaussian modeling indicates their line centroids shift from $6.553^{+0.005}_{-0.008}
\rm~keV$/$6.820^{+0.011}_{-0.017}\rm~keV$ in 2000 to $6.626\pm0.010\rm~keV$/$6.883\pm0.011\rm~keV$ in 2011.
And these two absorption lines are quite narrow and can only be marginally resolved.
The derived $3\sigma$ upper limit of their line widths ($\sigma_{\rm E}$) are $<9\rm~eV$ / $<8\rm~eV$ respectively for 2000, and $< 5\rm~eV$ / $<3\rm~eV$ for 2011.
We chose to fix the line width of each line to $\sigma_{\rm E}=1.9\rm~eV$ ($FWHM=4.4\rm~eV$), which is the instrumental spectral resolution of HEG at 6.7 keV.
Then for each epoch and for each of the two absorption features individually, we tested the reduction of best-fit $C$-stat for the 2 extra degrees of freedom (d.o.f) after introducing the Gaussian absorption whose line centroid and intensity are allowed to vary.
The best-fit $\Delta C$ is -8.25 / -2.56 for epoch 2000, and -6.11 / -8.86 for epoch 2011 with respect to the baseline continuum model.
It should be noted that F-test cannot be used to evaluate the significance of a line  \citep{2002ApJ...571..545P}.
Therefore, we also performed Monte Carlo simulations to evaluate the likelihood that these absorption lines arise purely from statistical fluctuations\citep{2010A&A...521A..57T,2021NatAs...5..928S}.
We generated $10^4$ simulated spectra with the best-fit baseline continuum model only (\texttt{pow}+\texttt{pion}(abs)$_{\rm torus}$+\texttt{refl}) and the corresponding exposure time for each epoch. We scanned over 5 -- 8 keV with a narrow Gaussian line ($\sigma_{\rm E}=0\rm~eV$) to search for potential absorption features with the minimum $\Delta C$ in each simulated spectrum.
The probability $p$ that a random fluctuation in the simulated continuum produces an absorption feature with a larger absolute $\Delta C$ than that of the observed absorption line is calculated for each line and each epoch. The quantity $1-p$ therefore represents how confident the line detection would be.
The resulting confidence levels (1-$p$) for each of two absorption lines is 99.6\% / 94.1\% for epoch 2000, and 94.0\% / 97.7\% for epoch 2011, respectively.

The separations between the two detected absorption line centroids are of the similar value ($\Delta E\sim0.26\rm~keV$) and remain almost unchanged within error range across the two epochs, both consistent with the separation between resonant Fe XXV K$\alpha$ and Fe XXVI Ly$\alpha$ lines.
This lends support to our identification.
We have simulated $10^4$ continuum spectra and scanned for minimum $C$-stat for each epoch, but this time adding two Gaussian absorption models simultaneously. The separation of their line centroids is fixed at the separation between Fe XXV and Fe XXVI lines ($\Delta E=0.26\rm~keV$). We counted the fraction of simulated spectra with greater absolute $\Delta C$ than the observed best fit after introducing two absorption lines.
The fraction indicates that the probability of such pair of absorption lines arising from pure statistical fluctuation is $\textless0.01\%$ for epoch 2000 and $0.03\%$ for epoch 2011.

An apparent deceleration in the redshifted velocities is found from $\sim6.7\times10^3\rm~km~s^{-1}$ to $\sim3.5\times10^3\rm~km~s^{-1}$ consistently for both of the two absorption lines in the two epochs over 11 yrs.
This lends support to the interpretation that the two absorption lines are not independent and should origin from the same dynamical system.
We have validated such deceleration may not arise from pure statistical fluctuations at a confidence level of nearly 90\%, although the possibility of apparent shifts caused by independent two sets of lines across epochs still cannot be fully ruled out.
A detailed discussion can be referred to the Appendix \ref{sec:line_shift}.

We searched for other possible pairs of strong transition lines that can co-exist and match the observed separation between line centroids for common elements aside from highly ionized iron (e.g. O, Ne, Mg, Si, S, Ar and Ca).
We do find O VIII Ly$\alpha$ (rest-frame 0.654 keV) and resonant Ne IX K$\alpha$ lines (rest-frame 0.922 keV) meet the requirement, but it may correspond to extremely large relativistic blueshifted velocity ($\sim0.98c$).

Thus, we believe the most probable identification is that the two absorption features are assigned to Fe XXV K$\alpha$ and Fe XXVI Ly$\alpha$ lines of the same absorber,
{\bf with confidence levels of $\sim94-99.6\%$ for both 2000 and 2011 epoch}.

\subsubsection{Inflowing Absorber Inside the Torus}\label{subsubsec:redWA_in_torus}
The absorption features are present in both epochs of spectra, spanning 11 years.
We replace the phenomenological Gaussian models with physical model, to compare the differences in the physical properties of the ionized absorbers.
We add an additional redshifted absorber \texttt{pion}(abs)$_{\rm red}$ to account for this highly ionized inflowing absorption component and place it inside the neutral torus absorber, closer to the central black hole.

\begin{figure}[hbtp!]
\centering\includegraphics[width=0.45\textwidth]{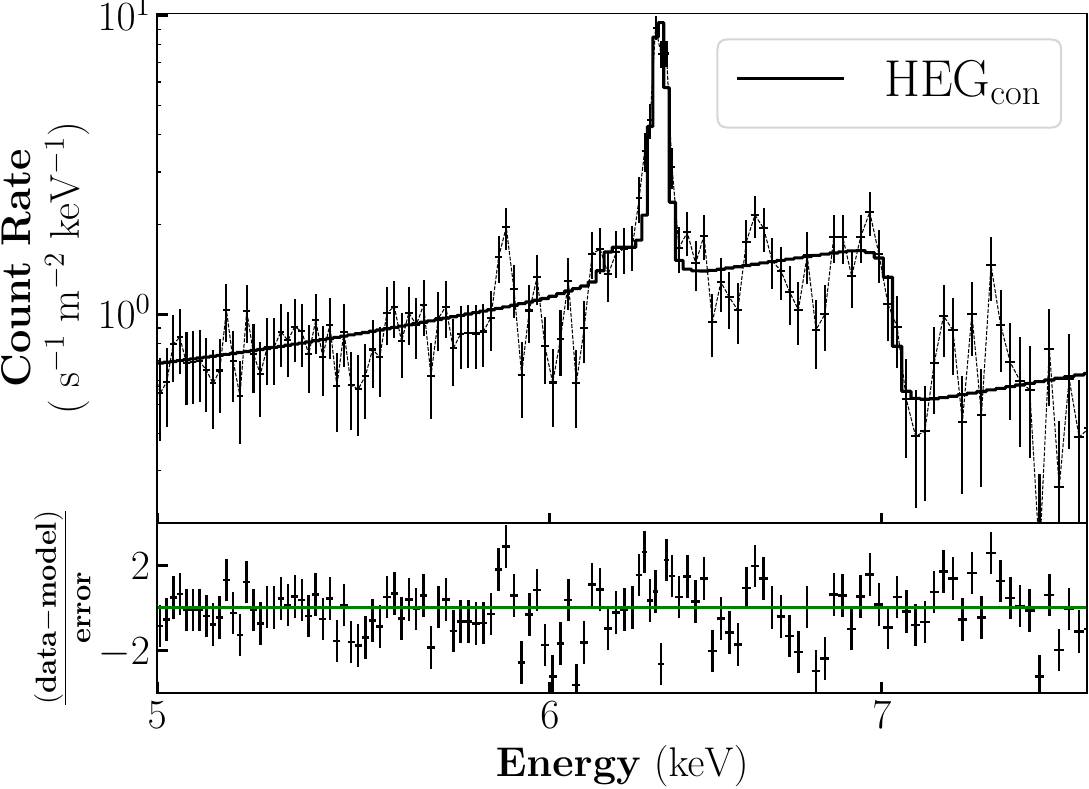}
\caption{Co-added HEG spectra of Mrk\,3. Black solid line represents the best-fit continuum model, involving a power-law continuum, a reflection component and neutral absorption from torus.
Residual absorption line features remain around $6.5$--$6.8\rm~keV$.}
\label{fig:heg_tot}
\end{figure}

The centroids of the two highly ionized Fe absorption lines consistently shift by $\Delta v\sim3\times10^3\rm~km~s^{-1}$ from 2000 to 2011, suggesting a potential deceleration of the ionized absorber between the two epochs.
To compare the differences in properties of the redshifted absorber across epochs, we fit the spectra of the two epochs separately.
The best-fit results are summarized in the columns of HEG$_{\rm 2000}$ and HEG$_{\rm 2011}$ of Table \ref{tab:best_fit_HEG}.

Both the photon index of the illuminating power-law and the column density of torus for the two epochs are roughly consistent with the baseline fitting of co-added spectrum (HEG$_{\rm con}$) in Sec. \ref{subsec:base}.
However, the reflection fraction ($\sim0.07$ and $\sim0.031$) becomes smaller.
This reduction is due to the presence of the additional redshifted absorber model, which attenuates the incident continuum and leads to a higher best-fit normalization of the unabsorbed power-law component $I_{\rm PL}$.
The actual absolute magnitude of the reflection component $I_{\rm refl}$ remains unchanged. As a result the reflection fraction $scal=\frac{I_{\rm refl}}{I_{\rm PL}}$ decreases.

Over the 11 years, the best-fit velocity of the putative inflowing absorber changes from $6.1\pm0.5\times10^3\rm~km~s^{-1}$ to $3.4\pm0.3\times10^3\rm~km~s^{-1}$,
consistent with the fitting results of Gaussian models in Section\ref{subsubsec:line_identi}.
Its velocity broadening $\sigma_{\rm v}$ also slightly increases from $\sim8\times10^2\rm~km~s^{-1}$ to $\sim2.7\times10^3\rm~km~s^{-1}$, with large uncertainties.
Both the column density and ionization parameter increase with the decreasing of the bulk velocity across epochs.

\begin{deluxetable*}{l|lc|c|cc|c}
\tablecaption{Best-fit results for HEG spectral modeling}
\scriptsize
\label{tab:best_fit_HEG}
\tablewidth{0pt}
\tablehead{
\colhead{Mod} & \colhead{Par} & \colhead{Unit} &  \colhead{${\rm HEG}_{\rm con}$} & \colhead{${\rm HEG}_{\rm 2000}$} & \colhead{${\rm HEG}_{\rm 2011}$} & \colhead{${\rm HEG}_{\rm alter}$} \\
\colhead{(1)} & \colhead{(2)} & \colhead{(3)} & \colhead{(4)} & \colhead{(5)} & \colhead{(6)} & \colhead{(7)}
}
\startdata
epoch &     &           & co-added & 2000 & 2011 & co-added\\
band &     & $\rm \AA$ & 1.6 -- 6.2 & 1.6 -- 6.2 & 1.6 -- 6.2 & 1.6 -- 6.2\\ \hline
\multirow{2}{*}{\texttt{pow}} & $\Gamma$ & - &  $2.24\pm0.06$ & $2.07_{-0.08}^{+0.07}$ & $2.32_{-0.06}^{+0.05}$ & $2.25\pm0.06$ \\ 
{    } & Norm & $\rm10^{52}~ph~s^{-1}~keV^{-1}$ & $8_{-2}^{+3}$ & $16_{-8}^{+19}$ & $60_{-17}^{+25}$ & $10_{-3}^{+4}$ \\ \hline
\texttt{refl} & scal & - & $0.20_{-0.05}^{+0.07}$ & $0.07_{-0.04}^{+0.07}$ & $0.031_{-0.009}^{+0.011}$ & $0.026_{-0.005}^{+0.005}$ \\ \hline
\multirow{2}{*}{\texttt{pion}(abs)$_{\rm torus}$} & ${\rm log}(\xi)$ & $\rm log(erg~cm~s^{-1})$ &  $-2$ (fixed) & $-2$ (fixed) & $-2$ (fixed) & $-2$ (fixed)\\
{    } & $N_{\rm H}$ & $\rm 10^{24}~cm^{-2}$ & $1.73_{-0.13}^{+0.14}$ & $2.1_{-0.3}^{+0.4}$ & $1.82_{-0.14}^{+0.16}$ &  $1.85_{-0.13}^{+0.15}$ \\ \hline
\multirow{4}{*}{\texttt{pion}(abs)$_{\rm red}$} & ${\rm log}(\xi)$ & $\rm log(erg~cm~s^{-1})$ &  - & $3.5_{-0.2}^{+0.6}$ & $4.5_{-0.3}^{+0.4}$ & - \\
{    } & $N_{\rm H}$ & $\rm 10^{24}~cm^{-2}$ &  - & $0.26_{-0.13}^{+1.01}$ & $2.0_{-1.8}^{+2.1}$ & - \\
{    }  & $v_{\rm bulk}$ & $\rm 10^3~km~s^{-1}$ & - & $+6.1\pm0.5$ & $3.4\pm0.3$ & -\\
{    } & $\sigma_{\rm v}$ & $\rm km~s^{-1}$ & - & $8_{-4}^{+5}\times10^2$ & $2.7_{-1.1}^{+1.3}\times10^2$ & - \\ \hline
\multirow{3}{*}{\texttt{hot}(abs)$_{\rm red}$} & $T$ & keV & - & - & - & $15_{-4}^{+9}$ \\
{  } & $N_{\rm H}$ & $\rm 10^{24}~cm^{-2}$ & - & - & - & $2.79\pm0.07$ \\
{  } & $v_{\rm bulk}$ & $\rm 10^3~km~s^{-1}$ & - & - & - & $3.3_{-0.2}^{+0.3}$  \\ \hline
$C$--stat &     &      & 849.32 & 663.39 & 777.94 & 832.37 \\
cAIC & & & 1076.8 & 1070.0 & 1105.8 & 1072.4 \\
d.o.f &     &      & 612 & 590 & 607 & 609
\enddata
\tablecomments{
\footnotesize
(1) Spectral epochs, energy bands and model components adopted for the spectral fitting. \texttt{pow}: the illuminating power law from the central source;
\texttt{refl}: the reflection component from the torus;
\texttt{pion}(abs): photoionized material absorbing the illuminating source, the subscript `torus' denotes an almost neutral absorber with ionization parameter log($\xi$) fixed at $-2$, while `red' represents a red-shifted photoionized absorber;
\texttt{hot}(abs)$_{\rm red}$: a red-shifted collisionally ionized absorber.
(2)--(3): Model parameters and units.
$\Gamma$: photon index;
Norm: normalization of the power law in units of $10^{52}\rm~ph~s^{-1}~keV^{-1}$.
scal: reflection fraction;
log($\xi$): 10-base log of ionization parameter in units of $\rm erg~cm~s^{-1}$;
$N_{\rm H}$: column density in units of $\rm 10^{24}~cm^{-2}$;
$T$: gas temperature in units of $\rm keV$;
$v_{\rm bulk}$: bulk motion velocity of the gas in units of $\rm 10^3~km~s^{-1}$;
$\sigma_{\rm v}$: velocity dispersion of the gas in units of $\rm km~s^{-1}$
(4) Continuum model for the co-added HEG spectrum, consisting of an illuminating power law absorbed by an almost neutral torus together with the torus reflection.
(5)--(6) Spectral fitting for 2000 and 2011 epochs (denoted by suffixes) respectively. The total model further adds the red-shifted photoionized absorber \texttt{pion}(abs)$_{\rm red}$ to the continuum model.
(7) An alternative model for the co-added spectrum, where a red-shifted, collisionally ionized absorber \texttt{hot}(abs)$_{\rm red}$ is added to the continuum model instead.
}
\end{deluxetable*}

\subsubsection{Alternative geometry}\label{subsubsec:alter_geo}
We also examined an alternative geometry, placing the redshifted absorber outside the torus.
However, this configuration fails to produce a good fit, as most of the high energy photons are already absorbed by the inner torus and cannot further ionize Fe ions to the observed high states.

Another possible scenario is that the highly ionized absorber is not inside the torus and photoionized by the central AGN, but instead represents hot, dense collisionally ionized gas in the host galaxy that is moving away from us along line of sight.
Based on the baseline continuum model in Sec. \ref{subsec:base}, a collisionally ionized absorber model \texttt{hot}(abs)$_{\rm red}$ is placed outside the torus to fit the co-added HEG spectrum, as shown in the column HEG$_{\rm alter}$ in Table \ref{tab:best_fit_HEG}.
This model gives a relative good fit with $C$-stat ($832.4$ for $609$ d.o.f).
The temperature of the putative hot gas reaches $15_{-4}^{+9}\rm~keV$.

\subsection{Modeling the Soft X-ray Band}\label{subsec:softX}

Multiple emission lines are clearly present in the co-added RGS spectrum over 6.2--24.6 $\rm\AA$ (0.5--2 keV) band.
Instead of phenomenological Gaussian models, we adopt physical emission models of collisionally ionized/photoionized gas embedded in SPEX (\texttt{CIE}/\texttt{PION}).
We adopt a similar underlying baseline continuum model as described in Section \ref{subsec:base}, consisting of an incident power-law continuum absorbed by the AGN torus. 
The reflection component is negligible in the 0.5--2 keV band, and cannot be well constrained from RGS spectrum alone without the energy coverage of neutral 6.4 keV Fe K$\alpha$ and Compton hump.
So we choose to ignore it for simplification.
The ionization parameter $\xi$ and column density $N_{\rm H}$ of the torus are fixed at the values derived from co-added HEG spectrum.

\begin{figure*}[hbtp!]
\centering\includegraphics[width=0.95\textwidth]{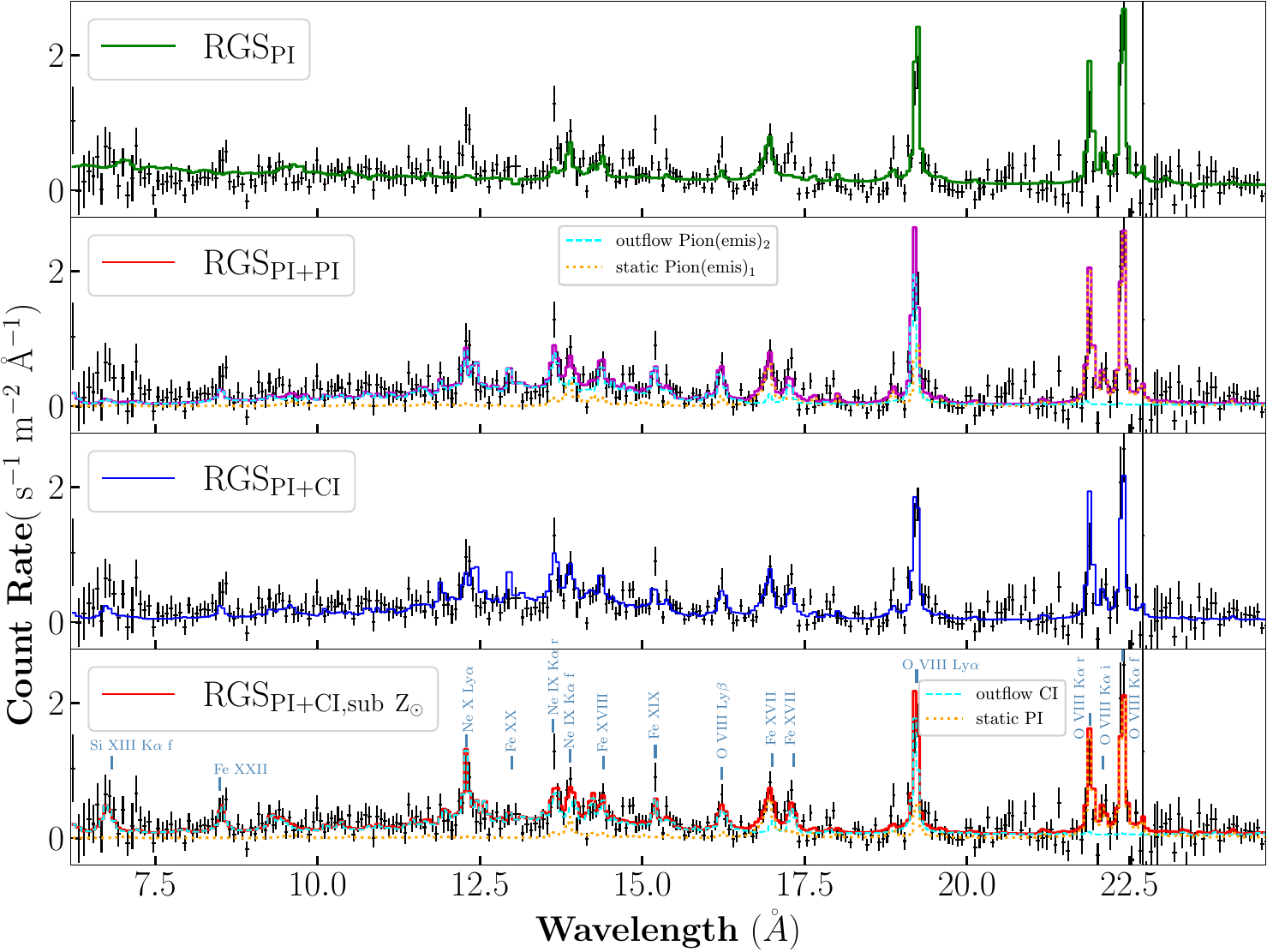}
\caption{
Spectral modeling of co-added RGS spectrum for Mrk\,3 over $6.2$ -- $24.8\rm~\AA$. Black crosses mark the observed data and error. Solid lines represent the best-fit total models.
The baseline continuum consists of an incident power-law, a reflection component and torus absorption fixed at the best-fit value from co-added HEG spectrum.
{\it The first panel}: Spectrum fitted with one single photoionized emission gas component (\texttt{pion}(emis)$_{\rm 1}$) in addition to the baseline. Clear residual lines remain around $8.5\rm~\AA$, $12.5\rm~\AA$ and $17.2\rm~\AA$.
{\it The second panel}: Adding the second photoionized gas component (\texttt{pion}(emis)$_{\rm 2}$, marked with cyan dashed line) would overestimate the continuum level of short wavelength $\lesssim10\rm~\AA$.
It has higher ionization parameter and dominates the shorter wavelength.
The first, static photoionized gas component (\texttt{pion}(emis)$_{\rm 2}$, marked as orange dotted line) with lower ionization parameter dominates the longer wavelength around O VII triplets.
{\it The third panel}: Substituting the second \texttt{pion}(emis)$_{\rm 2}$ with a collisionally ionized gas component (\texttt{cie}) in solar metallicity.
{\it The forth panel}: The most preferred model consists of a static photoionized gas (orange dotted line) and an outflowing collisionally ionized gas with sub-solar iron abundance $\sim0.22\rm~Z_{\odot}$ (cyan dashed line). The temperature of the gas is $\sim 1\rm~keV$ and the blueshifted velocity is $\sim740~\rm km~s^{-1}$.
}
\label{fig:rgs}
\end{figure*}

First, we try to reproduce the multiple lines with emission from a single static photoionized gas component (\texttt{pion}(emis)$_{\rm 1}$).
We denote this model set as ${\rm RGS}_{\rm PI}$.
Gas in the vicinity of the black hole can naturally be illuminated and ionized by the central AGN.
We assume that the power-law continuum irradiating the static photoionized gas is identical to that irradiating the torus.
The covering factor $omeg$ of \texttt{pion}(emis)$_{\rm 1}$ is fixed at unity for simplicity.
The best-fit column density $N_{\rm H}$ is $2.9\pm0.3\times10^{21}\rm~cm^{-2}$.
Detailed fitting results can be refer to Table \ref{tab:best_fit_RGS} and Figure \ref{fig:rgs}.
The best-fit model underestimates the emission lines around $8.5\rm~\AA$, $12.5\rm~\AA$ and $17.2\rm~\AA$.

Adding a second photoionized emission component (\texttt{pion}(emis)$_{\rm 2}$) improves the fit with a $C$-stat reduction of in $119.5$ for $5$ extra degrees of freedom (d.o.f), with a F-test probability as low as $1.5\times10^{-13}$.
Detailed best-fit parameters of this new model set are listed in the column $\rm RGS_{PI+PI}$ in Table \ref{tab:best_fit_RGS}.
The illuminating power-law continuum and the bulk motion velocity of this new \texttt{pion}(emis)$_{\rm 2}$ are allowed to vary, representing a potential (narrow-line region) outflow responding to a different illuminating continuum to the static photoionized gas.
And the best-fit $\Gamma$ for the outflowing photoionized gas is $1.26\pm0.06$, roughly consistent with the fiducial model set $\rm RGS_{PI+PI}$.
As can be seen from the second panel of Figure \ref{fig:rgs}, the static photoionized gas (\texttt{pion}(emis)$_{\rm 1}$) with lower ionization parameter dominates the longer wavelength around O VII triplets, while \texttt{pion}(emis)$_{\rm 2}$ outflowing at a velocity of $\sim6.7_{-0.8}^{+0.9}\times10^2\rm~km~s^{-1}$ and with higher ionization parameter, is responsible for emission lines in shorter wavelength below $\sim19\rm~\AA$, e.g. Ne X, Fe XVII and Fe XVIII.
This model slightly overpredicts the continuum level at shorter wavelength $\lesssim7.5\rm~\AA$.
Both the static and outflowing photoionized gas metallicity are assumed to have solar abundance. We allowed the iron abundance to vary freely, but the best-fit abundance still remains around $1$. 
A more flexible model is examined, in which the illuminating power law for the torus, the static, and the outflowing photoionized components are all allowed to independently vary. However,
the fit cannot converge successfully under this assumption.
We have also tried an alternative assumption that the torus and the outflowing \texttt{pion}(emis)$_{\rm 2}$ share the same illuminating power law. This results in a slightly worse $C$-stat of $571$.

We attempt to substitute the outflowing photoionized component \texttt{pion}(emis)$_{\rm 2}$ with a collisionally ionized component with solar metallicity.
The best-fit result is recorded in the column $\rm RGS_{\rm PI+CI}$ of the Table \ref{tab:best_fit_RGS} and Figure \ref{fig:rgs}.
The temperature of the \texttt{cie} gas is $0.96\pm0.02\rm~keV$
and it is outflowing with a blueshifted velocity of $\sim3.3\pm0.6\times10^2\rm~km~s^{-1}$.
Compared to RGS$_{\rm PI+PI}$, this model reduces $C$-stat for $12$ with even one fewer free parameter, potentially implying a better fit.
However, F-test can only be used if the new model is nested within the old one \citep{2002ApJ...571..545P} and is not valid in this case.
To further assess the goodness of fit, we calculate the corrected Akaike Information Criterion ($cAIC=-2lnL+2k+\frac{2k^2+2k}{n-k-1}$) \citep{1974ITAC...19..716A,2002JApSt..29..267F,2007MNRAS.377L..74L}, where $k$ represents the number of free model parameters and $n$ denotes the number of spectral bins. The likelihood $L$ between the model $m_{\rm i}$ and data $d_{\rm i}$ is described by $L=\prod\limits_{\rm i}\frac{exp(-m_{\rm i})m_{\rm i}^{d_{\rm i}}}{d_{\rm i}!}$ for $C$-stat.
The cAIC does not require the model to be nested, making it particularly suitable for evaluating alternative physical scenarios.
And compared with standard AIC, cAIC is more strict by introducing an extra term of $\frac{2k^2+2k}{n-k-1}$ which strengthens the penalty of extra free parameters for small number of spectral bins \citep{burnham2002model}.
The cAIC for RGS$_{\rm PI+CI}$ ($321.0$) is lower by 7 compared to RGS$_{\rm PI+PI}$ ($328.0$).
According to \citet{2007MNRAS.377L..74L}, $\Delta \rm cAIC>5$ would lead to a strong claim that one model is better than the other.
This implies that $\rm RGS_{\rm PI+CI}$ is favored. But compared with $\rm RGS_{\rm PI}$ (cAIC$=319.0$), adding the extra collisionally ionized gas increases cAIC up to 2.0 for 3 extra free parameters.

Also, there remain some discrepancies between model and observation around Si XIII and Fe XXII lines, probably due to enhancement in element abundance.
Then we allow the iron abundance of the outflowing CIE gas to vary freely. A new best-fit summarized in Table \ref{tab:best_fit_RGS} with the column of $\rm RGS_{\rm PI+CI,sub~Z_{\odot}}$ has been achieved, with a sub-solar iron abundance $Z_{\rm Fe}=0.22\pm0.03\rm~Z_{\odot}$.
Compared to the result with solar metallicity (RGS$_{\rm PI+CI}$), the C-stat reduces significantly of 82.7 for one extra free parameter
, and the cAIC for RGS$_{\rm PI+CI,sub~Z_{\odot}}$ ($313.6$) further reduces $7.4$ for one extra free parameter.
This corresponds to a strong claim that the collisional gas with sub-solar iron abundance is favored.

Adding further one more photoionized or collisional ionized component does not improve the goodness of fit and thus is not necessary.
So we conclude that the PI+CI with sub-solar metallicity scenario has the lowest cAIC and C-stat, and is the preferred model.
We examined the effect of ignoring reflection component (\texttt{refl}) to this scenario. After adding a reflection component with parameters derived from HEG spectral fitting for epoch 2000 and 2011 respectively, we found only a negligible reduction in $C$-stat ($-0.03$ and $-0.01$).
This confirms that neglecting the reflection component during RGS fitting is valid.

\begin{deluxetable*}{l|lc|cccc}
\tablecaption{Best-fit results for RGS spectral modeling}
\tiny
\label{tab:best_fit_RGS}
\tablewidth{0pt}
\tablehead{
\colhead{Mod} & \colhead{Par} & \colhead{Unit} & \colhead{${\rm RGS}_{\rm PI}$} & \colhead{${\rm RGS}_{\rm PI+CI}$} & \colhead{${\rm RGS}_{\rm PI+PI}$} & \colhead{RGS$_{\rm PI+CI,sub~Z_{\odot}}$} \\
\colhead{(1)} & \colhead{(2)} & \colhead{(3)} & \colhead{(4)} & \colhead{(5)} & \colhead{(6)} & \colhead{(7)}
}
\startdata
band &     & $\rm \AA$ & 6.2 -- 24.8 & 6.2 -- 24.8 & 6.2 -- 24.8 & 6.2 -- 24.8 \\ \hline
\multirow{2}{*}{\texttt{pow}} & $\Gamma$ & - & $1.006_{-0.014}^{+0.009}$ & $2.1\pm0.5$ & $2.4\pm0.4$ & $2.9^{+0.5}_{-0.7}$\\ 
{    } & Norm & $\rm10^{50}~ph~s^{-1}~keV^{-1}$ & $6.1\pm0.4$ & $4.8_{-0.9}^{+1.1}$ & $4.7_{-0.8}^{+1.0}$ & $2.7_{-0.8}^{+0.9}$\\ \hline
\multirow{2}{*}{\texttt{pion}(abs)$_{\rm torus}$} & ${\rm log}(\xi)$ & $\rm log(erg~cm~s^{-1})$ &  $-2$ (fixed) &  $-2$ (fixed) &  $-2$ (fixed) & $-2$ (fixed)\\
{    } & $N_{\rm H}$ & $\rm 10^{24}~cm^{-2}$ & $1.85$ (fixed) & $1.85$ (fixed) & $1.85$ (fixed) & $1.85$ (fixed) \\ \hline
\multirow{2}{*}{\texttt{pion}(emis)$_{\rm 1}$} & ${\rm log}(\xi)$ & $\rm log(erg~cm~s^{-1})$ & $1.48\pm0.02$ & $0.97_{-0.06}^{+0.36}$ & $0.99_{-0.17}^{+0.35}$ & $1.3^{+0.7}_{-0.5}$ \\
{    } & $N_{\rm H}$ & $\rm 10^{21}~cm^{-2}$ & $2.9\pm0.3$ & $2.7_{-0.6}^{+0.8}$ & $2.9\pm0.3$ & $3.6^{+1.9}_{-1.0}$\\ \hline
\multirow{3}{*}{\texttt{cie}} & T & $\rm keV$ &  - & $0.96\pm0.02$ & - & $0.86\pm0.03$\\
{    } & $v_{\rm bulk}$ & $\rm 10^2~km~s^{-1}$ & - & $7.4_{-0.9}^{+0.5}$ & - & $3.3\pm0.6$\\
{    } & $Z_{\rm Fe}$ & $\rm Z_{\odot}$ & - & 1 (fixed) & - & $0.22\pm0.03$ \\
{    } & Norm & $\rm 10^{64}~cm^{-3}$ & - & $1.20\pm0.05$ & - & $2.56\pm0.17$ \\ \hline
\multirow{4}{*}{\texttt{pion}(emis)$_{\rm 2}$} & ${\rm log}(\xi)$ & $\rm log(erg~cm~s^{-1})$ & - & - & $2.10\pm0.02$ & - \\
{    } & $N_{\rm H}$ & $\rm 10^{21}~cm^{-2}$ & - & - & $0.404_{-0.008}^{+0.026}$ & - \\
{    }  & $v_{\rm bulk}$ & $\rm 10^2~km~s^{-1}$ & - & - & $6.7_{-0.8}^{+0.9}$ & - \\
{  } & $\Gamma_{\rm incident}$ &  & - & - & $1.11\pm0.02$ & - \\ \hline
$C$--stat &     &     & 699.41 & 568.01 & 579.86 & 485.35\\
cAIC &  &  & 319.0 & 321.0 & 328.0 & 313.6 \\
d.o.f &     &      & 304 & 301 & 299 & 300
\enddata
\tablecomments{
\footnotesize
(1) Energy bands and model components adopted for the spectral fitting.
\texttt{pow}: the illuminating powerlaw from the central source;
\texttt{pion}(abs)$_{\rm torus}$: the almost neutral torus, with ionization parameter log($\xi$) fixed at $-2$;
\texttt{pion}(emis): emission from photoionized gas outside the line of sight. The subscript `1' and `2' denotes the first and second emission component;
\texttt{cie}: emission from collisionally ionized gas.
(2)-(3): Model parameters and units. $Z_{\rm Fe}$ denotes the iron abundance relative to solar value; $\Gamma_{\rm incident}$ represents the photon index of the powerlaw illuminating the second photoionized emitting component. Definitions of the remaining parameters follow those in Table \ref{tab:best_fit_HEG}.
(4) RGS$_{\rm PI}$: The model comprises an illuminating powerlaw continuum (\texttt{pow}) absorbed by neutral torus (\texttt{pion}(abs)$_{\rm torus}$), and one photoionized emitter (\texttt{pion}(emis)$_{\rm 1}$) illuminated by the same powerlaw.
(5) RGS$_{\rm PI+CI}$: an additional collisionally ionized gas (\texttt{CIE}) is added to RGS$_{\rm PI}$.
(6) RGS$_{\rm PI+PI}$: a second photoionized emitter (\texttt{pion}(emis)$_{\rm 2}$) has been added to RGS$_{\rm PI}$. This extra component is assumed to be illuminated by a powerlaw distinct from \texttt{pow}, with photon index $\Gamma_{\rm incident}$ free to vary.
(7) RGS$_{\rm PI+CI,sub~Z_{\rm \odot}}$: same as RGS$_{\rm PI+CI}$, but the iron abundance of collisionally ionized gas is allowed to vary. This scenario is preferred.
}
\end{deluxetable*}

\section{Discussion} \label{sec:diss}
In the following section, we discuss the origin, kinematics, possible deceleration mechanism and the interaction with gas environment of the redshifted absorber near the central AGN in Mrk\,3. 

For the 1.6--6.2 $\rm\AA$ hard X-ray spectra observed with Chandra HEG, the best-fit continuum model consists of an illuminating power-law, an almost neutral torus absorption and reflection component, along with a redshifted photoionized absorber characterized by two redshifted absorption lines. This redshifted absorber exhibits clear variations in its physical properties between epoch 2000 and epoch 2011. In the following analysis, we use the best-fit parameters listed in the column HEG$_{\rm 2000}$ and HEG$_{\rm 2011}$ of Table \ref{tab:best_fit_HEG} to estimate the physical and dynamical properties of the redshifted absorber, including its bulk velocity ($v_{\rm bulk}$), ionization parameter ($\xi$), and column density ($N_{\rm H}$).

For the co-added XMM-Newton RGS spectra covering the 6.2--24.6 $\rm\AA$ soft X-ray band, the continuum is accompanied by multiple prominent emission lines. These lines can be described by a nearly static photoionized emission gas and an outflowing collisionally ionized gas component with sub-solar iron abundance. We adopt the ionization parameter $\xi$ and column density $N_{\rm H}$ of the \texttt{pion}(emis)$_{\rm 1}$, as well as the temperature and normalization of the \texttt{cie} listed in column RGS$_{\rm PI+CI,sub~Z_{\odot}}$ of Table \ref{tab:best_fit_RGS} to further characterize the physical conditions of the gas environment around the central AGN.

These preferred best-fit spectral models are selected based on the criteria of minimizing both cAIC and C-stat.

\subsection{Inflowing torus clumps}\label{subsec:inflow_clump}
A redshifted, highly ionized absorber at a high velocity of several $10^3\rm~km~s^{-1}$ has been detected from the HEG spectrum, {\rm apparently} persistent over a timescale of at least $11\rm~yrs$.
The most probable explanation for this inflowing while decelerating highly ionized absorber is that, 
instability at the outer surface of the torus leads to the evaporation of the torus clumps.
These clumps are stretched and drawn inward toward the accretion flow under the gravitational influence of the central SMBH, and subsequently become highly ionized under the intense radiation field, as schematically illustrated in Figure \ref{fig:sketch}.
Due to the clumpy nature of the torus suggested by \citet{2016MNRAS.460.1954G},
this process is expected to occur stochastically yet persistently, involving multiple clumps over time.
The observed H-like and He-like iron absorption lines in different epochs thus may trace a continuous inflowing process of one clump or different clumps that connects between the clumpy torus and accretion disk.
Here we assume that these infalling torus clumps share  similar sizes and typical physical properties, and they could initiate infalling from similar distance (i.e., the torus).
This enables us to treat them equivalently as a single infalling clump at different stages.
In the following sections, we discuss the kinematics and interactions between these clumps and the surrounding medium at different distance to the central black hole based on this assumption, although the possibility that observed absorption features originate from {\bf distinct} transient events across epochs cannot be fully ruled out.

\subsubsection{Location of the inflowing absorber}
\label{subsubsec:RA_loc}
Assuming that the thickness of this absorber is comparable to its distance from the ionizing source (i.e, $\Delta r=r\lesssim\frac{L_{\rm ion}}{N_{\rm H}\xi}$), we can estimate its number density and maximum radial location for 2000 and 2011 epochs, following standard practice in previous warm absorber studies \citep{2001ApJ...561..684K,2016MNRAS.457.3896L,2022A&A...657A..77W}.
The ionization luminosity of Mrk\,3 over $13.6\rm~eV$ -- $13.6~\rm keV$ is computed directly by \texttt{SPEX} based on the best-fit illuminating power-law.
The redshifted absorber is located at a distance of $r_{\rm max}\lesssim0.74_{-0.53}^{+1.60}\rm~pc$ ($\sim3.4\times10^4\rm~r_{g}$) since 2000, and reduces to $\lesssim0.04_{-0.03}^{+0.10}\rm~pc$ ($\sim1.8\times10^3\rm~r_{g}$) in 2011 considering the 1$\sigma$ error range, consistent with the observed increase in ionization parameter over the time.
As the absorber approaching the central black hole, its number density increases by nearly two orders of magnitude, while the line-of-sight velocity drops by $\sim2.7\times10^{3}\rm~km~s^{-1}$.
The derived properties of the infalling ionized absorber at both epochs are recorded in Table \ref{tab:prop_RA}.

\begin{deluxetable}{l|cccc}[hbtp!]
\tablecaption{Properties of the infalling absorber}
\scriptsize
\label{tab:prop_RA}
\tablewidth{0pt}
\tablehead{
\colhead{Epoch} & \colhead{$L_{\rm ion}$} & \colhead{$r_{\rm max}$} & \colhead{$n_{\rm H}$} & \colhead{$M_{\rm abs}/C_{\rm f,abs}$}\\
\colhead{(1)} & \colhead{(2)} & \colhead{(3)} & \colhead{(4)} & \colhead{(5)}
}
\startdata
2000 & $1.9^{+2.3}_{-1.0}$ & $0.74_{-0.53}^{+1.60}$ & $1.1_{-0.8}^{+2.6}\times10^5$ & $4.7_{-0.5}^{+3.3}\times10^4$\\
2011 & $8.6^{+3.6}_{-2.4}$ & $0.04_{-0.03}^{+0.10}$ & $1.5_{-1.1}^{+3.7}\times10^7$ & $1.1_{-0.7}^{+3.5}\times10^3$\\
\enddata
\tablecomments{
\footnotesize
The 1$\sigma$ uncertainties are calculated through bootstrapping.
(1) Spectral epoch;
(2) The ionizing luminosity over $13.6\rm~eV$--$13.6\rm~keV$ in units of $\rm 10^{45}~erg~s^{-1}$, and is directly computed from the best-fit illuminating power-law component by \texttt{SPEX};
(3) Distance of the redshifted absorber in units of $\rm pc$, assuming its depth $\Delta r$ is equivalent to the distance $r$;
(4) Estimated number density of the redshifted absorber in units of $\rm cm^{-3}$.
(5) Estimated total mass over covering factor of the redshifted absorber, in units of $\rm M_{\odot}$. 
}
\end{deluxetable}

The redshifted absorption features exist in two epochs spanning $11~\rm yrs$. By multiplying this timescale with the inflowing velocity, we estimate the dynamical length (or depth) of the absorber along line-of-sight to be $r_{\rm dyn}\sim0.04-0.07\rm~pc$.
This value is comparable to the inferred $r_{\rm max}$ in 2011, lending support to our previous assumption that $\Delta r=r$.

The dust sublimation radius of Mrk\,3 is estimated to be $R_{\rm sub}=0.5L_{\rm 46}^{0.5}(\frac{\rm 1800}{T_{\rm sub}}^{2.6}f(\theta))\sim0.13\rm~pc$, where $L_{\rm 46}$ is AGN bolometric luminosity in units of $10^{46}\rm~erg~s^{-1}$ \citep{1987ApJ...320..537B}.
For Mrk\,3, the unabsorbed $2$ -- $10\rm~keV$ luminosity from our best-fit model is $L_{\rm 2-10}=5.1\times10^{44}\rm~erg~s^{-1}$, corresponding to an intrinsic bolometric luminosity of $L_{\rm bol}\sim5.1\times10^{45}\rm~erg~s^{-1}$ after applying a bolometric correction factor $10$ \citep{2020A&A...636A..73D}.
For simplification, the dust sublimation temperature $T_{\rm sub}$ is chosen to be $1800\rm~K$ following \citet{2016MNRAS.457.3896L} and the angular-dependent geometrical factor is fixed at 1.
This suggests the detected inflowing ionized absorber lies at a distance either interior to or comparable with the dusty torus of Mrk\,3,
consistent with our spectral modeling assumption that the inflowing absorber resides closer to the SMBH than the torus.

\subsubsection{Mass of the absorber and the torus}
\citet{2016MNRAS.460.1954G} reported a variability of
$5$ -- $10$\% in the line-of-sight column density ($N_{\rm H}$) within $3$ -- $30\rm~d$ based on SED fitting of Mrk\,3 NuSTAR spectra.
These occultation events were interpreted as transits of small cloudlets within the torus, providing strong support for the clumpy nature of Mrk\,3 torus.
Notably, the best-fit column density of the highly-ionized redshifted absorber ($N_{\rm H,red}$) in our analysis is of the same order of magnitude ( $\sim10^{24}\rm~cm^{-2}$) as that of the neutral torus ($N_{\rm H,torus}$).
Furthermore, the characteristic number density of individual torus clumps would be $\sim10^5$ -- $10^7\rm~cm^{-3}$ \citep{Netzer_2013}, comparable to the estimated number density we infer for the infalling absorber ($1.1\times10^5$ -- $1.5\times10^7\rm~cm^{-3}$).
These consistencies add weight to our aforementioned scenario in which the redshifted absorber originates from torus clumps.

To compare the total mass of the observed infalling absorber $M_{\rm abs}$ with that of the AGN torus $M_{\rm torus}$,
we assume both of them are homogeneous.
Each of these structures is approximated as partial sections of a spherical shell surrounding the central AGN. That is, the material occupies a fraction of the shell with inner ($R_{\rm in}$) and outer radii ($R_{\rm out}$) corresponding to the inner and outer boundaries of the torus or the absorber. Thus the total mass $M$ is then proportional to the volume of this shell $M=\mu m_{\rm P} n_{\rm H}\frac{4}{3}\pi(R_{\rm out}^3-R_{\rm in}^3)*C_{\rm f}$, scaled by a covering factor
$C_{\rm f}$, which can be interpreted as a combination of sky coverage ($\frac{\Omega}{4\pi}$) and volume filling factor inside the absorber or torus.
For the absorber $M_{\rm abs}$, we take $R_{\rm in}=r_{\rm max}$ estimated in Section \ref{subsec:inflow_clump} and Table \ref{tab:prop_RA} and assume the depth of the absorber is equivalent to its distance $R_{\rm out }=R_{\rm in}+\Delta r\sim2R_{\rm in}\sim2r_{\rm max}$. The total mass of the infalling ionized absorber decreases from $M_{\rm abs,2000}\sim4.7_{-0.5}^{+3.3}\times10^{4}\rm~M_{\odot}*C_{\rm f,abs}$ in epoch 2000 to $M_{\rm abs,2011}\sim1.1_{-0.7}^{+3.5}\times10^{3}\rm~M_{\odot}*C_{\rm f,abs}$ in epoch 2011, recorded in Table \ref{tab:prop_RA}.
For the AGN torus, we adopt a characteristic number density ($n_{\rm H,torus}\sim10^{4-5}\rm~cm^{-3}$) and the best-fit average column density $N_{\rm H,torus}\sim2\times10^{24}\rm~cm^{-2}$ from our spectral modeling. The inner radius of the torus $R_{\rm in,torus}$ is assumed to correspond to the dust sublimation radius $R_{\rm sub}\sim 0.13\rm~pc$.
The total mass of AGN torus is estimated to be $M_{\rm torus}\sim5\times10^{(5-7)}\rm~M_{\odot}*C_{\rm f,torus}$.
While the covering factors, $C_{\rm f,abs}$ and $C_{\rm f,torus}$ are hard to constrain,
it is reasonable to assume that the torus has larger or comparable covering factor as the redshifted absorber.
So the estimated torus mass is of several orders of magnitude larger than the ionized absorber, consistent with our scenario that the infalling absorber originates from the torus.

\subsection{Kinematics of the inflowing absorber}
We compare the mass inflow rate of the detected ionized absorber with the intrinsic accretion rate of Mrk\,3.
The mass inflow rate of the redshifted ionized absorber is estimated as $\dot{M}_{\rm in,WA}=4\pi\mu m_{\rm P}n_{\rm H}v_{\rm in}r_{\rm max}^2\times C_{\rm f}\sim(36-163~\rm M_{\odot}~yr^{-1})\times C_{\rm f}${, where $v_{\rm in}$ is the best-fit red-shifted velocity of the inflowing absorber from HEG spectra of epoch 2000 and 2011.}
Homogeneous absorber is assumed for simplification and a covering factor $C_{\rm f}$ is included.
The intrinsic bolometric luminosity of Mrk\,3 is estimated to be $\sim5.1\times10^{45}\rm~erg~s^{-1}$ as discussed in Section \ref{subsec:inflow_clump}
Assuming a $\eta=10\%$ energy transfer efficiency, the mass accretion rate at the event horizon is $\dot{M}_{\rm in,EH}\equiv\frac{L_{\rm bol}}{\eta {\rm c}^2}\sim0.9\rm~M_{\odot}~yr^{-1}$.

The inflow rate of the ionized absorber is almost 50 times higher than the mass accretion rate at the event horizon.
These infalling torus clumps are sufficient to supply the accretion and maintain the nuclear activity in Mrk\,3, provided that the correction factor $C_{\rm f}\gtrsim0.03$.
Probably these evaporated torus clumps possess a small sky covering factor above the torus surface
and form an episodic but ongoing inflow,
fueling the accretion of central AGN. This may also explain the rarity of redshifted absorber detections, which requires a grazing line of sight to torus.
Alternatively, and more likely, only $\sim(0.6-3)$\% of the inflowing material finally reaches the accretion disk, with the majority being entrained and ejected by a putative disk wind.
The derived $n_{\rm H}$ and $N_{\rm H}$ of the infalling ionized absorber increase during its deceleration,
while its velocity dispersion $\sigma_{\rm v}$ decreases, implying that the velocity gradient is being suppressed.
These observational signatures further supports the scenario that the ionized absorber is compressed and becomes denser as it approaches the SMBH.

\begin{figure}[hbtp!]
\centering\includegraphics[width=0.45\textwidth]{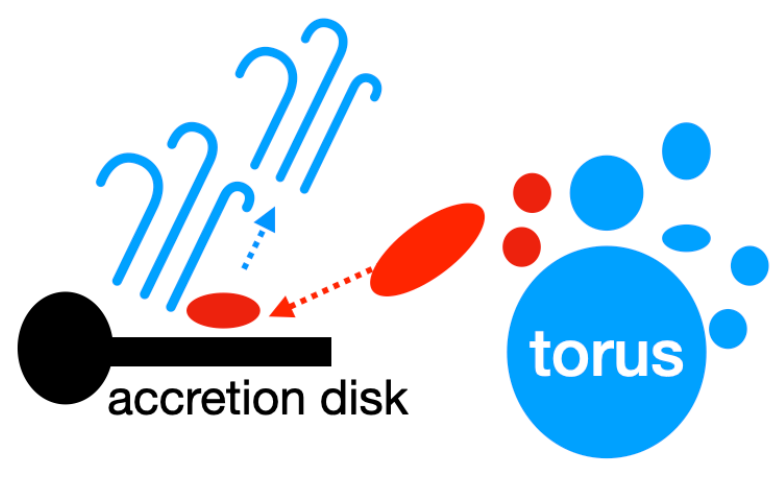}
\caption{
A diagram for the sub-parsec scale gas environment around Mrk\,3. Some torus clumps are evaporated and fall toward the central black hole, serving as a disk-feeding inflow. The inflowing ionized absorber would be decelerated or even ejected by a putative disk wind.
\label{fig:sketch}}
\end{figure}

\subsection{Shocked interstellar medium in the NLR}
\label{subsection:NLR_shock}
The soft X-ray emission of Mrk\,3 can be described by one photoionized plasma component together with one outflowing collisionally ionized plasma component.

The photoionized component is likely associated with emission from the narrow-line region and is nearly static, with a $3\sigma$ upper limit of $\lesssim173\rm~km~s^{-1}$ for outflow velocity.
In the optical band, three photoionized gas components have been adopted to describe emission lines detected by Hubble {\it STIS} in the NLR of Mrk\,3 \citep{2005ApJ...619..116C, 2009ApJ...694..765C}.
In their studies, the ionization parameters of the NLR gas components are reported in $U$, defined as $U\equiv\frac{Q}{4\pi r^2n_{\rm e}c}$. $Q$ is the number of ionizing photons, $r$ represents the distance to the illuminating source, $n_{\rm e}$ is the electron number density, and $c$ is the speed of light.
To compare with their results, we convert our best-fit $\xi$ of the photoionized emission gas \texttt{pion}(emis)$_{\rm 1}$ to $U$. We took advantage of the fact that SPEX self-consistently fits both the photoionized gas and the ionizing power-law simultaneously. The value of $U$ corresponding to our best-fit $\xi$ can be read from \texttt{SPEX} directly.
We find that the ionization parameter of our soft X-ray photoionized plasma ($\rm{log}(U)\sim-0.8$) is even higher than that of the most highly ionized optical components (${\rm log}(U)\sim-1.5$) reported,
suggesting that the soft X-ray trace emission from the inner part of NLR, close to the central AGN.
Considering a typical gas density of NLR to be $\sim10^3$ -- $10^5\rm~cm^{-3}$, the soft X-ray emission is estimated to come from gas around $0.8$ -- $8\rm~pc$, based on the best-fit ionization parameter. 

The collisionally ionized plasma component, with a blueshifted velocity of $\sim330\rm~km~s^{-1}$, could trace the shocked interstellar medium (ISM) as proposed by \citep{2017ApJ...848...61B}.
They suggest that a shocked gas component extends toward east and west up to $\sim278\rm~pc$ from Chandra imaging observations, in alignment with the extended biconical warm ionized gas outflows traced by optical emission lines (e.g., [O III] and H$\alpha$). Based on the density ratio between the pre-shock and post-shock region, they derived the shock velocity is $\sim600-1000\rm~km~s^{-1}$.
Our best-fit outflowing \texttt{cie} gas temperature ($\sim0.86\rm~keV$) and iron abundance ($\sim0.22\rm~Z_{\odot}$) from RGS spectrum are consistent with those derived from their MEG observation ($\sim0.85\rm~keV$ and $Z=0.24$).
And the blueshifted velocity of our detected soft X-ray outflow is roughly in agreement with
the average velocity of a co-spatial [O III] outflow ( $\sim500\rm~km~s^{-1}$), which also spans over $10$ -- $300\rm~pc$ \citep{2001AJ....122.2961R, 2020ApJ...893...80G, 2023ApJ...943...98M}.
This may imply that the soft X-ray shocked gas and the [O III] may share the same origin.
The outflow rate of [O III] gas peaks at $\sim8\rm~M_{\odot}~yr^{-1}$ around $200\rm~pc$ from the SMBH \citep{2021ApJ...910..139R}, indicating that the biconical shocked outflow in the narrow line region may be generated or accelerated at larger distances, potentially undergoing interaction with a radio jet or putative disk wind from the central black hole.

\subsection{Possible deceleration mechanism}
If we assume the observed apparent deceleration of red-shifted absorption features comes from a single infalling clump within 11 years, this would
suggest the presence of an extra force acting against the gravitational pull of the central SMBH.
The observed apparent deceleration rate is $a_{\rm obs}\sim-0.73\rm~cm~s^{-2}$ with the negative sign denoting force direction away from SMBH).
The outward radiative acceleration exerted on the redshifted absorber ($a_{\rm rad}\equiv\frac{F_{\rm abs}}{\mu m_{p}N_{H}c}\sim7\times10^{-15}\rm~cm~s^{-2}$) is significantly smaller than the inward gravitational acceleration $a_{\rm grav}\equiv\frac{GM_{\rm BH}}{r^2}\sim0.01-3.92\rm~cm~s^{-2}$.
$F_{\rm abs}$ represents the absorbed flux and
we take the solar proton-to-electron abundance $\mu=1.4$.
The black hole mass $M_{\rm BH}=4.5\times10^8\rm~M_{\rm\odot}$ is taken from \citet{2002ApJ...579..530W} based on stellar velocity dispersion. The distance to the black hole $r$ is taken to be the derived location $r_{\rm max}$ of the absorber for each epoch (See Section \ref{subsec:inflow_clump} and Table \ref{tab:prop_RA}).
The radiation pressure is insufficient to compete with gravity,
likely because the observed ionized absorber is too close to the central illuminating source and is already heavily ionized.
This is consistent with the results of \citet{2007MNRAS.374..237L}, where they adopted a simplified model to describe the velocity structure of gas under the gravitational pull of central BH while competing with radiation pressure exert on electrons. Their model predicts that pure radiative pressure cannot compete with gravitational pull, and the gas would accelerate monotonically as its distance to BH decreases. In contrast, our spectral analysis reveals that the infalling absorber is decelerating toward BH, indicating that extra deceleration mechanism is required.

One possible explanation for the deceleration is ram pressure $P_{\rm ram}$ exerted by surrounding medium.
The deceleration exerted on the inflowing absorber by such a wind can be estimated as $a_{\rm ram}\equiv\frac{P_{\rm ram}*S_{\rm abs}}{M_{\rm abs}}=-\frac{\rho (v_{\rm wind}+v_{\rm abs})^2 S_{\rm abs}}{M_{\rm abs}}$, where $S_{\rm abs}$ stands for the cross-sectional area of between infalling absorber and the wind.
We also assume the covering factor of the infalling absorber is equivalent to its sky coverage $C_{\rm f,abs}=\frac{\Omega}{4\pi}=\frac{S_{\rm abs}}{4\pi r_{\rm max}^2}$ and $r_{\rm max}$ is the derived location of the absorber.
Thus the ram pressure deceleration from the putative wind can be expressed as $a_{\rm ram}=-\mu m_{\rm p}n_{\rm wind}(v_{\rm wind}+v_{\rm abs})^2 4\pi r_{\rm max}^2\frac{C_{\rm f,abs}}{M_{\rm abs}}$. Here the relative speed between the putative wind and inflowing absorber $v_{\rm wind}+v_{\rm abs}$ is considered.
The velocity of the absorber $v_{\rm abs}$ is taken from the best-fit value of the inflowing photoionized absorber velocity $v_{\rm bulk}$ recorded in the columns HEG$_{\rm 2000}$ and HEG$_{\rm 2011}$ of Table \ref{tab:best_fit_HEG}, which is $\sim6.1\times10^3\rm~km~s^{-1}$ for epoch 2000 and $\sim3.4\times10^3\rm~km~s^{-1}$ for epoch 2011.
The total mass scaled with covering factor $\frac{M_{\rm abs}}{C_{\rm f,abs}}$ and the estimated location $r_{\rm max}$ of the absorber for each epoch from Table \ref{tab:prop_RA}.
$v_{\rm wind}$ and $n_{\rm wind}$ represents the velocity and number density of the putative outflow.

A collisionally ionized outflow has been detected from RGS spectrum of Mrk\,3.
We take the best-fit outflow velocity $v\sim330\rm~km~s^{-1}$ from the preferred model RGS$_{\rm PI+CI,sub~Z_{\rm \odot}}$ in Table \ref{tab:best_fit_RGS} as $v_{\rm wind}$.
The density $n_{\rm wind}$ of this collisionally ionized outflow is hard to constrain without the knowledge of volume occupation $V_{\rm wind}$. Here we naively assume it homogeneously and spherically fills up to the derived location of the infalling absorber and $V_{\rm wind}\sim\frac{4}{3}\pi r_{\rm max}^3$. Based on the best-fit emission measure $Norm\sim n_{wind}^2V_{\rm wind}$ recorded in Table \ref{tab:best_fit_RGS}, we are able to estimate the $n_{\rm wind}\sim2.3\times10^4-1.8\times10^6\rm~cm^{-3}$.
Such an RGS detected soft X-ray outflow could provide a decelerating rate of $a_{\rm ram,RGS}\sim-0.016\rm~cm~s^{-2}$ to $-0.05\rm~cm~s^{-2}$.
The boundary of the range represents the estimation for epoch 2000 and 2011 respectively.
The predicted deceleration rate is about one order of magnitude smaller than the observed value.
Considering the large uncertainties involved in the estimation, the 0.86-keV hot outflow detected in the RGS spectrum as one of the potential candidates for braking the redshifted absorber is not favored, but cannot be fully ruled out. 

As discussed in Section \ref{subsection:NLR_shock}, this RGS detected collisionally ionized outflow is more likely associated with extended shocked NLR gas, the several hundreds of $\rm km~s^{-1}$ velocity of which is too low to be the initially launched disk wind.
Highly ionized, high-velocity outflows are prevalent in luminous AGN. Ultra-fast outflows (UFOs), characterized by blueshifted K-shell absorption lines of H-like and He-like Fe with velocity exceeding $10^{4}\rm~km~s^{-1}$ can be found in up to $40$\% in local Seyfert galaxies \citep{2014MNRAS.443L.104T}.
A UFO with a typical number density $n_{\rm wind}$ around $\sim(2-4)\times10^{6}\rm~cm^{-3}$ and a velocity of $\sim10^4\rm~km~s^{-1}$ can provide a deceleration rate $a_{\rm ram,UFO}\sim-3.7$--$-0.5\rm~cm~s^{-2}$, comparable to the observed value.
However, such a disk wind is elusive to detection, probably due to the narrow opening angle of such UFO and the misalignment between its outflow direction and our line of sight toward Mrk\,3.
While the NLR outflow detected in both soft X-ray and optical [O III] band may be driven by such putative disk wind and serve as an indirect evidence.

We have also considered the ram pressure braking by the dense broad line region (BLR) clouds.
Similarly, the deceleration caused by the ambient, nearly static BLR clouds ($v_{\rm wind}\sim0\rm~km~s^{-1}$) can be expressed as $a_{\rm ram, BLR}=-\mu m_{\rm p}n_{\rm BLR}v_{\rm abs}^2 4\pi r_{\rm max}^2\frac{C_{\rm f,abs}}{M_{\rm abs}}$.
Here $v_{\rm abs}$ describes the relative speed between the inflowing absorber and the dense BLR clouds.
Assuming a typical average density for the almost static BLR clouds $n_{\rm BLR}\sim10^8\rm~cm^{-3}$,
the deceleration rate is estimated to be $-2.6$ -- $-61\rm~cm~s^{-2}$, which is also sufficient, possibly even excessive, to brake the fast-infalling material.
Given the clumpy nature of BLR gas, such braking effect may not occur continuously.
We speculate that the BLR cloud is also a plausible candidate to cause the observed deceleration.

\subsection{Coherent scenario}
Overall, we propose a coherent scenario in Mrk\,3 as illustrated by Figure \ref{fig:sketch}.
The redshifted, highly-ionized absorber observed in the hard X-ray band traces the inflowing clumps originated from torus instability.
These clumps are evaporated from the torus at sub-parsec scales and serve as potential fuel for black hole accretion, falling roughly along the opening angle of the torus surface.
Mrk\,3, with a grazing line-of-sight configuration, may represent a local analog to the small sample of $z\sim1$ quasars, where parsec-scale inflows connecting torus and disk are characterized by broad/narrow optical absorption lines \citep{2019Natur.573...83Z, 2022A&A...659A.103C}.
The apparent deceleration of the inflowing ionized absorber may indirectly support the existence of a putative disk wind.
Such a disk wind could also entrain and eject the accretion material away, explaining the reason why only a small fraction of the inflowing material finally reaches the black hole event horizon.
The disk wind can propagate outward and interact with the surrounding ISM, traced by shocked collisionally ionized $\sim1\rm~keV$ gas in soft X-ray band.
It is also a promising candidate for driving or accelerating the biconical [O III] outflows in the NLR up to hundred-parsec scale.
However, there are no clear spectral signatures of such a disk wind, probably because the opening angle of the wind is narrow and its outflow direction is out of our sight line.
So other possible deceleration mechanisms including ram pressure BLR clouds cannot be fully ruled out.

\subsection{Other possibilities} 
There is possibility that the observed redshifted absorption features are caused by gravitational redshift instead of the Doppler effect of the infalling motion. Based on the location $r_{\rm max}$ of the absorber derived from photoionization modeling (See Section \ref{subsubsec:RA_loc}), we estimated the potential gravitational redshift $z_{\rm grav}\equiv(1-\frac{2r_{\rm g}}{r_{\rm max}})^{-\frac{1}{2}}-1$ to be $\sim3\times10^{-5}$ for epoch 2000 and $5\times10^{-4}$ for epoch 2011, assuming the central black hole is non rotating.
These values are significantly smaller than the observed redshifts, making this scenario unlikely in interpreting the apparent velocity.

A series of studies suggests a `failed wind' scenario that an initially launched disk wind close enough to the AGN would become over ionized by the radiation field as it moving outward, until the radiation driving force cannot support the wind cloud to further escape. Such gas cloud would fall back toward the accretion disk, forming a redshifted `failed wind'\citep{2003ApJ...582...69P,2004ApJ...616..688P}.
Similarly, \citet{2004A&A...413..535G} proposed the `aborted jet' scenario that if the central engine works intermittently, previously ejected blobs of material would fall back.
Both of these scenarios could manifest as redshifted absorber and cannot be fully ruled out with current observations.

\section{Summary} \label{sec:sum}
We analyze the high resolution X-ray spectroscopic data of Mrk\,3 nuclei and provide the direct evidence of sub-parsec scale feeding inflows bridging the gap between torus and SMBH accretion disk. 
Our main results are as follows:

\begin{itemize}
\item A fast inflowing, highly ionized absorber characterized by redshifted Fe XXV and Fe XXVI absorption lines has been detected in the Chandra HEG spectra.
It decelerates from $6.1\pm0.5\times10^3\rm~km~s^{-1}$ to $3.4\pm0.3\times10^3\rm~km~s^{-1}$ over 11 years.
Photoionization modeling suggests that both the ionization parameter and column density increase during the deceleration, accompanied with a decrease in the distance to central black hole from $\lesssim0.74\rm~pc$ to $\lesssim0.04\rm~pc$.

\item The inflowing ionized absorber is estimated to be located between the outer accretion disk and torus, with a mass inflow rate of $\sim(36-163)\times C_{\rm f}\rm~M_{\odot}~yr^{-1}$. Compared with the accretion rate of Mrk\,3, only
$\sim(0.6-3)$\% of the inflowing material finally reaches the event horizon.

\item Multiple emission lines detected in soft X-ray band can be well described by an almost static gas component illuminated by the central AGN, together with a collisionally ionized gas component with temperature of $0.86\pm0.03\rm~keV$ and sub solar metallicity ($Z_{\rm Fe}\sim0.22\rm~Z_{\odot}$) outflowing at a velocity of $\sim330\rm~km~s^{-1}$.

\item We propose a probable scenario to explain the gas dynamics in sub-parsec scale in Mrk\,3.
Torus clumps are evaporated and fall toward central black hole due to instability, manifesting as redshifted ionized absorber in a grazing line of sight to torus.
The inflowing material is decelerated by the ram pressure from either dense BLR cloud or a putative ultra-fast disk wind. Such a disk wind lacks direct detection but is supported by the existence of a soft X-ray outflowing gas and the biconical [O III] outflow in the NLR of Mrk\,3.

\end{itemize}

AGN with such grazing-view in the local Universe like Mrk\,3 would enable us to pierce into the fueling inflow connecting torus and outer accretion disk.
With high-resolution X-ray spectroscopy like Chandra HETG and XRISM Resolve, we would expect to find more cases with similar orientation. 
A comprehensive study of both inflow and outflow down to sub-parsec scale near the central engine is vital to obtain a clear insight of the feedback effect connecting multi-scale, multi-phase gas environment around SMBH.

\begin{acknowledgments}
We thank Junjie Mao, Daniele Rogantini, Mengfei Zhang, Yulong Gao, Suoqing Ji, Feng Yuan and Zhiyuan Li for helpful discussions and critical comments.
We are also grateful for the anonymous referee to help us improve the manuscript a lot.
We would like to acknowledge the organizing committee of the {\it I-HOW COSPAR Workshop 2024: A New Era of High-Resolution X-ray Spectroscopy} who inspired the basic idea and provided a platform for collaboration of this work.
F.S. is supported in part by the China Postdoctoral Science Foundation (grants 2022TQ0354 and 2022M723279).
Y.J.W. acknowledges support by National Natural Science Foundation of China (Project No. 12403019) and Jiangsu Natural Science Foundation (Project No. BK20241188).
\end{acknowledgments}

%

\vspace{5mm}
\facilities{CXO, XMM-Newton}


\software{CIAO\citep{2006SPIE.6270E..1VF}, SAS\citep{2004ASPC..314..759G}, 
SPEX\citep{kaastra_2024_spex},
PION\citep{2016A&A...596A..65M}
          }

\appendix
\section{Shift of line centroids across epochs}
\label{sec:line_shift}
\begin{figure}[hbtp!]
\centering\includegraphics[width=0.96\textwidth]{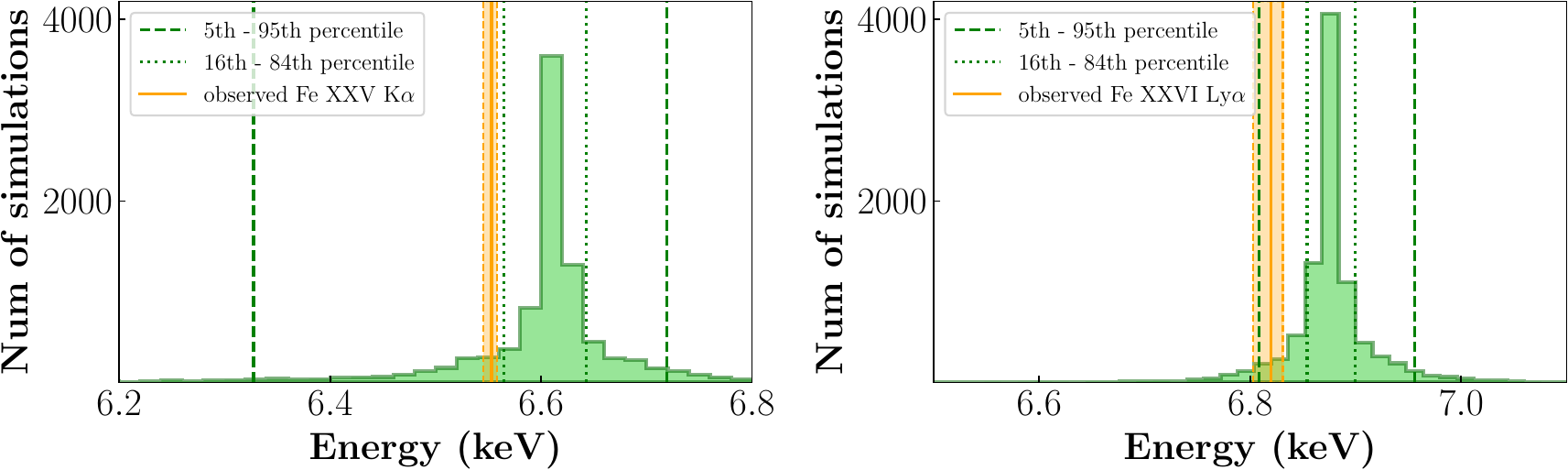}
\caption{
Distributions of line centroids obtained from $10^4$ simulated spectra based on 2011 best-fit model to match the lower flux and signal-to-noise ratio of 2000 observation for Fe XXV K$\alpha$ line (left panel) and Fe XXVI Ly$\alpha$ line (right panel).
The dashed / dotted green vertical lines mark the $5^{\rm th}$-$95^{\rm th}$ / $16^{\rm th}$-$84^{\rm th}$ percentiles of line centroids from simulated spectra. The orange strips represent the observed line centroids (solid orange line) and 1-$\sigma$ error ranges for epoch 2000. For both absorption features, the observed values lie at least outside the $16^{\rm th}$-$84^{\rm th}$ percentiles of expected ranges.}
\label{fig:MC_lineshift}
\end{figure}

We performed Monte Carlo simulations to validate that the detected deceleration of the absorption line centroids are real and do not arise from statistical fluctuations across the epoch 2000 and 2011, similar to the methods described in \citet{2025arXiv250507963P}.
Since the continuum level and signal-to-noise ratio of epoch 2000 are lower than epoch 2011, 
we generated $10^4$ simulated spectra with the best-fit continuum model of epoch 2000 together with the Gaussian line models from epoch 2011. The simulated exposure time is chosen to be the same as epoch 2000.
For each simulated spectrum, we fit with the continuum model described in Sec.\ref{subsec:base} plus two gaussian absorption models, and the two line centroids are recorded.
Their distributions are shown in Figure \ref{fig:MC_lineshift}. The $5^{\rm th}$, $16^{\rm th}$, $84^{\rm th}$ and $95^{\rm th}$-percentiles of simulated line centroids are marked and compared with the real observed values with 1-$\sigma$ error ranges of epoch 2000.
For both two lines the observed values lie outside the $16^{\rm th}$-percentiles, and Fe XXVI Ly$\alpha$ line is even close to $5^{\rm th}$-percentile.
More importantly, both absorption lines showing centroid shifts of consistent magnitude and direction.
Therefore, this supports the interpretation that the two absorption line centroids consistently shift across the epochs at a confidence level of nearly 90\%,
although the possibility of independent lines still cannot be fully ruled out.

\bibliography{sample631}{}
\bibliographystyle{aasjournal}


\end{CJK*}
\end{document}